\documentclass[twocolumn,english,superscriptaddress,prl]{revtex4}
\usepackage{amssymb}
\usepackage{subfigure}
\usepackage{amsmath}
\usepackage{graphicx}
\usepackage{xcolor}

\begin{document}

\title{Independent Control of Scattering Lengths in Multicomponent Quantum
Gases}
\author{Peng Zhang$^{1}$, Pascal Naidon$^{1}$ and Masahito Ueda}
\affiliation{ERATO, JST, Macroscopic Quantum Control Project, Hongo, Bunkyo-Ku, Tokyo
113-8656, Japan }
\affiliation{Department of Physics, University of Tokyo, Hongo, Bunkyo-Ku, Tokyo
113-8656, Japan }

\begin{abstract}
We develop a method of simultaneous and independent control of
different
scattering lengths in ultracold multicomponent atomic gases, such as $^{%
\mathrm{40}}$K or $^{\mathrm{40}} $K-$^{\mathrm{6}}$Li mixture. Our
method can be used to engineer multi-component quantum phases and
Efimov trimer states.
\end{abstract}

\maketitle

\textit{Introduction.} Recently, multicomponent gases of degenerate fermions
\cite{fermiontheory,fermionexp,pascal} or boson-fermion mixtures \cite{bf}
have attracted broad interest both theoretically and experimentally. Novel
quantum phases \cite{fermiontheory} and Efimov states \cite{efimov,pascal}
have been predicted in three-component Fermi gases. If interspecies
scattering lengths can be altered independently, one can engineer Efimov
states and the quantum phases of a three-component Fermi gas, and control an
effective interaction between two-component fermions immersed in a Bose gas
\cite{bf}. The magnetic Feshbach resonance (MFR) technique \cite%
{MFTR} can control only one scattering length, or a few scattering
lengths but not independently. The optical Feshbach resonance
technique \cite{OFR} can be directly generalized to the independent
control of more than one scattering length \cite{zhang}. However, it
would significantly shorten the lifetime of the system unless
the atom has a long-lived electronic excited state (e.g., the $^3\mathrm{P}%
_{0}$ state of Yb atom \cite{Takahashi}).

In this Letter, we combine the two ideas of MFR and rf-field-induced
Feshbach resonance \cite{rfscattering3} to propose a method for
independently controlling two scattering lengths in three-component
atomic gases. In our scheme, atoms are dressed via couplings between
different hyperfine states. With a proper magnetic field, the
independent tuning of scattering lengths of atoms in different
dressed states can be achieved by individual control of the Rabi
frequencies and detunings for different couplings.

\begin{figure}[tbp]
\includegraphics[bb=59bp 119bp 582bp 676bp,clip,width=8cm]{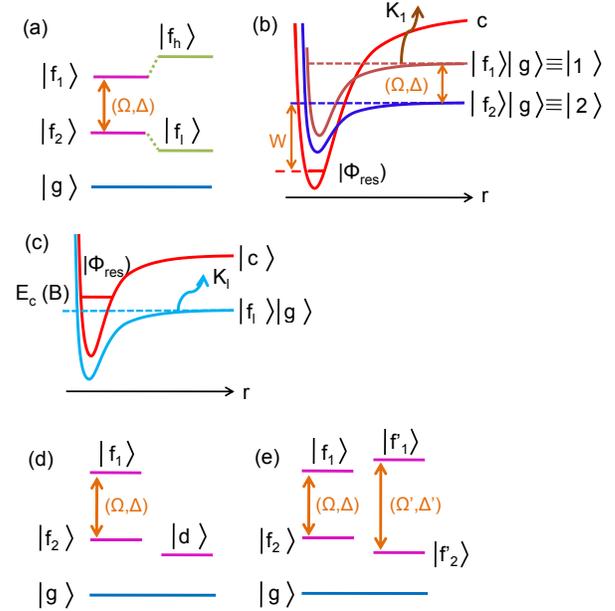}
\caption{(color online) (a) Schematic hyperfine levels used for the
control of a single scattering length. The states $|f_{1,2}\rangle$
are coupled to form two dressed states $|f_{h,l}\rangle$. (b) Bare
channels with hyperfine states
$|1\rangle\equiv|f_{1}\rangle|g\rangle$ and
$|2\rangle\equiv|f_{2}\rangle|g\rangle$ are coupled with parameters
$(\Omega,\Delta)$, and $|2\rangle$ is coupled to the bound state
$|\phi_{\mathrm{res}})$ in the closed channel with hyperfine state
$|c\rangle$ via interaction $W$. The state $|1\rangle$ can decay via
a HFR process with two-body loss rate $K_1$. (c) Dressed channels.
$a_{lg}$
is resonantly enhanced when the threshold energy of the channel with $%
|f_l\rangle|g\rangle$ crosses the bound state $|\Phi_{\rm res})$
in $%
|c\rangle$. (d), (e) Hyperfine states used for the control of two
scattering lengths with our first (d) and second (e) methods. All
the levels are plotted in the rotating frame of reference. In (b)
and (c), $r$ indicates the interatomic distance.}
\end{figure}

\textit{Control of a single scattering length.}  We first consider
atoms with
three ground hyperfine levels (Fig. 1a) $|f_1\rangle$, $|f_2\rangle$, and $%
|g\rangle$. In this Letter, we use the Dirac bracket $|\rangle$ to
denote the hyperfine levels of one or two atoms; $|)$ to indicate
the spatial states of the relative motion between two atoms, and
$|\rangle\rangle$ for the total two-atom state, which includes both
spatial motion and hyperfine state. We assume that $|f_1\rangle$ is
coupled to $|f_2\rangle$ through Rabi frequency $\Omega$ and
detuning $\Delta$. Such a coupling can be realized with a two-color
stimulated Raman process (TCSRP), i.e., coupling the hyperfine
states $|f_{1,2}\rangle$ via a common excited state $|e\rangle$ (not
shown in the figure) by two laser beams \cite{tcsrp}. The atomic
loss caused by the spontaneous decay of $|e\rangle$ can be
suppressed when the laser frequencies are far detuned from the
resonance. We assume the Rabi frequencies to be at
least one order of magnitude smaller than the typical depth ($\sim$ a few $100$%
MHz) of optical traps (see e.g., \cite{lik}), so that the loss due
to the decay of the state $|e\rangle$ may be ignored. A stable
coupling can also be realized via an rf field that induces a direct
transition between two
hyperfine states. The Rabi frequency caused by the rf field can be $0.1$MHz$%
- $$1$MHz with the oscillating amplitude of magnetic field $\lesssim
0.1$G.

Without loss of generality, we assume that the scattering channel with respect to hyperfine state$%
|2\rangle\equiv|f_{2}\rangle |g\rangle $ is stable, while the channel with $%
|1\rangle\equiv|f_{1}\rangle |g\rangle $ can decay to another
channel with hyperfine state $|a\rangle$ through a hyperfine relaxation (HFR) process. %
We further assume that a static magnetic field is tuned to the
region of MFR
between the diatomic channel with $|2\rangle $ and a bound state $|\phi _{\mathrm{%
res}})$ in a closed diatomic channel with hyperfine state $|c\rangle
$ (Fig. 1b). The binding energy $E_{c}(B)$ of $|\phi
_{\mathrm{res}})$ can be controlled by magnetic field $B$.
Therefore, in the rotating frame of
reference, the Hamiltonian for the relative motion of two atoms is given by%
\begin{equation}
\hat{H}=\hat{H}^{(\mathrm{bg})}+E_{c}(B)|\phi _{\mathrm{res}}) (\phi
_{\mathrm{res}}|\otimes |c\rangle \langle c|+\hat{W}+\hat{W}^{\dag
}, \label{h1}
\end{equation}%
where%
\begin{eqnarray*}
\hat{H}^{(\mathrm{bg})} &=&-\nabla ^{2}\sum_{i=1,2,a}|i\rangle \langle i|+%
\hat{V}^{(\mathrm{bg)}}; \\
\hat{V}^{(\mathrm{bg)}} &=&\sum_{i=1,2,a}\left[ V_{i}^{(\mathrm{bg)}%
}(r)+E_{i}\right] |i\rangle \langle i| \\
&&+\left[ \Omega |1\rangle \langle 2|+V_{1a}(r)|1\rangle \langle a|+{\rm h.c.}%
\right] ; \\
\hat{W} &=&W(r)|\phi _{\mathrm{res}})(\phi _{\mathrm{res}%
}|\otimes |2\rangle \langle c|,
\end{eqnarray*}%
where we set $\hbar =2m_{\ast }=1$ with $m_{\ast }$ being the reduced mass. %
In Eq. (1), $V_{i}^{(\mathrm{bg)}}(r)$ ($i=1,2,a$) is the background
scattering potential in the channel $|i\rangle \ $with $r$ the
distance between the two atoms; $W(r)$ is the coupling between
$|2\rangle $ and $|c\rangle $; $V_{1a}(r)$ is the coupling between
$|a\rangle $ and $|1\rangle $; $E_{i}$ is the asymptotic
energy of the channel $|i\rangle $ in the rotating frame. Here we choose $%
E_{2}=0$, which implies $E_{1}=\Delta $ and $E_{a}=\Delta -\delta $ with $%
\delta $ being the energy gap between $|1\rangle $ and $|a\rangle $.

Hamiltonian (\ref{h1}) shows that scattering channels $|1\rangle $ and $%
|2\rangle $ are coupled via the Rabi frequency $\Omega $. Since this
coupling is given by the single-atom TCSRP, it does not vanish in the limit $%
r\rightarrow \infty $. The scattering length is therefore not well
defined for the bare channels $|1,2\rangle $. To overcome this
problem, we diagonalize the Hamiltonian by introducing dressed states $%
|f_{l}\rangle =\alpha |f_{1}\rangle +\beta |f_{2}\rangle $ and $%
|f_{h}\rangle =\beta |f_{1}\rangle -\alpha |f_{2}\rangle $ with eigenvalues $%
\mathcal{E}_{h/l}=\Delta /2\pm (\Omega ^{2}+\Delta ^{2}/4)^{1/2}$ and
coefficients $\alpha =\Omega \lbrack \Omega ^{2}+(\mathcal{E}_{l}-\Delta
)^{2}]^{-1/2}$ and $\beta =(\mathcal{E}_{l}-\Delta )[\Omega ^{2}+(\mathcal{E}%
_{l}-\Delta )^{2}]^{-1/2}$ which can be controlled via $\Delta $ and $\Omega
$. Since the effective coupling between the dressed scattering channels $%
|f_{l}\rangle |g\rangle $ and$|f_{h}\rangle |g\rangle $ vanishes in
the limit $r\rightarrow \infty $, the scattering length $a_{lg}$
between $|f_{l}\rangle $ and $|g\rangle $ is well defined.

In presence of the inter-channel coupling $\hat{W}$, both
$|f_{l}\rangle |g\rangle $ and $|f_{h}\rangle |g\rangle $ are
coupled with the bound state $|\phi _{\mathrm{res}})$ in the closed
channel $|c\rangle $. When the threshold $E_{l}$ of the channel $%
|f_{l}\rangle |g\rangle $ crosses the energy $E_{c}$ of $|\phi _{\mathrm{%
res}})$, the Feshbach resonance between $|f_{l}\rangle |g\rangle $
and $|\phi _{\mathrm{res}}) $ strongly alters the scattering length
$a_{lg}$ between $|f_{l}\rangle $ and $|g\rangle $. Therefore, for a
given magnetic field, one can control
$a_{lg}$ by tuning $E_{l}$ through the coupling parameters $(\Omega ,\ \Delta )$%
. By a straightforward generalization of the method in Ref.
\cite{julian}, we obtain
\begin{equation}
a_{lg}=a_{lg}^{(\mathrm{bg)}}-2\pi ^{2}\frac{\Lambda _{ll}-\kappa e^{2i\eta
}\Lambda _{al}}{\mathcal{D}-i\left( 2\pi ^{2}\right) \chi ^{1/2}\Lambda _{aa}%
}  \label{rea}
\end{equation}%
with the $(\Omega ,\Delta ,B)$-dependent parameters%
\begin{eqnarray*}
\mathcal{D} &=&E_{c}+\langle c|(\phi _{\mathrm{res}}|\hat{W}^{\dag
}G_{\mathrm{bg}}^{(P)}\hat{W}|\phi _{\mathrm{res}}) |c\rangle -%
\mathcal{E}_{l}; \\
\Lambda _{\alpha \beta } &=&\langle\langle \Phi _{\mathrm{bg}}^{(\alpha )}[0]|\hat{W%
}\hat{W}^{\dag }|\Phi _{\mathrm{bg}}^{(\beta )}[0]\rangle\rangle \
(\alpha ,\beta
=l,a); \\
\Gamma  &=&e^{2i\eta }\langle\langle \Phi _{\mathrm{bg}}^{(a)}[0]|\hat{W}\hat{W}%
^{\dag }|\Phi _{\mathrm{bg}}^{(l)}[0]\rangle\rangle ; \\
\kappa  &=&2\sqrt{\mathrm{Im}[a_{lg}^{(\mathrm{bg)}}]\chi}.
\end{eqnarray*}%
Here $a_{lg}^{(\mathrm{bg)}}$ is the background scattering length
between $|f_{l}\rangle $ and $|g\rangle $ in the absence of MFR with $|\phi _{%
\mathrm{res}})$, and $\chi\equiv\sqrt{\mathcal{E}_{l}-E_{a}}$;
$|\Phi _{\mathrm{bg}}^{(l,a)}[0]\rangle\rangle$ are the background
zero-energy scattering states with incident particles in the
channels $|f_{l}\rangle |g\rangle $ and $|a\rangle
$, that is, we have $|\Phi _{\mathrm{bg}}^{(l)}[0]\rangle\rangle=(1+G_{\mathrm{bg}}^{(+)}\hat{V}%
^{(\mathrm{bg)}})(2\pi )^{-3/2}|f_{l}\rangle |g\rangle $ and $%
|\Phi _{\mathrm{bg}}^{(a)}[0]\rangle\rangle=(1+G_{\mathrm{bg}}^{(+)}\hat{V}^{(%
\mathrm{bg)}})|\chi\hat{e}_z)|a\rangle $ with the zero-energy
background Green's functions $G_{\mathrm{bg}}^{(\pm )}=(i0^{\pm
}-\hat{H}^{(\mathrm{bg)}})^{-1}$ and $G_{\mathrm{bg}}^{(P)}=(G_{\mathrm{bg}%
}^{(+)}+G_{\mathrm{bg}}^{(-)})/2$;  $|\chi\hat{e}_z)$ satisfying
$(\vec{r}|\chi\hat{e}_z)=(2\pi )^{-3/2}e^{i\chi z}$ is the
eigenstate of the relative momentum with eigenvalue $\chi\hat{e}_z$,
and $\eta $ is determined by $\hat{V}^{\mathrm{(bg)}}.$

Equation (\ref{rea}) shows that one can control the effective
interaction between atoms in the states $|f_{l}\rangle $ and
$|g\rangle $, which is determined by the real part
$\mathrm{Re}[a_{lg}]$ of $a_{lg}$. Under a given magnetic field, one
can tune $\mathrm{Re}[a_{lg}]$ by changing the coupling parameters
$\left( \Omega ,\Delta \right) $ around the resonance point where
$\mathcal{D}=0$.

Due to the coupling term $V_{1a}$ in the Hamiltonian (\ref{h1}), the
HFR process also occurs from the dressed channel $|f_{l}\rangle
|g\rangle $ to $|a\rangle .$ The two-body loss rate $K_l$ due to
this HFR is proportional to the imaginary part $\mathrm{Im}[a_{lg}]$
of $a_{lg}$: $K_l=-16\pi \mathrm{Im}[a_{lg}]$ \cite{RMPFR}. In Eq.
(\ref{rea}), the change of $\mathrm{Im}[a_{lg}]$ due to $(\Omega
,\Delta )$ is described by $i\left( 2\pi ^{2}\right) \delta ^{1/2}\Lambda _{aa}$ and $i\mathrm{Im}%
[\kappa e^{2i\eta }\Lambda _{al}]$ . Due to these two terms, the
two-body loss is enhanced in the resonance region. To avoid this
difficulty, one should either use the scattering channels without
hyperfine relaxation for both $|f_{1}\rangle |g\rangle $ and
$|f_{2}\rangle |g\rangle $, or choose the proper atomic species for
which the parameters $\left( \left\vert \Lambda _{al}\right\vert
,\left\vert \Lambda _{aa}\right\vert ,\kappa \right) $ are
sufficiently small, so that the peak of $\mathrm{Im}[a_{lg}]$ is
much narrower than the resonance of $\mathrm{Re}[a_{lg}].$

\begin{figure}[tbp]
\includegraphics[bb=25bp 177bp 590bp 632bp,clip,width=9cm]{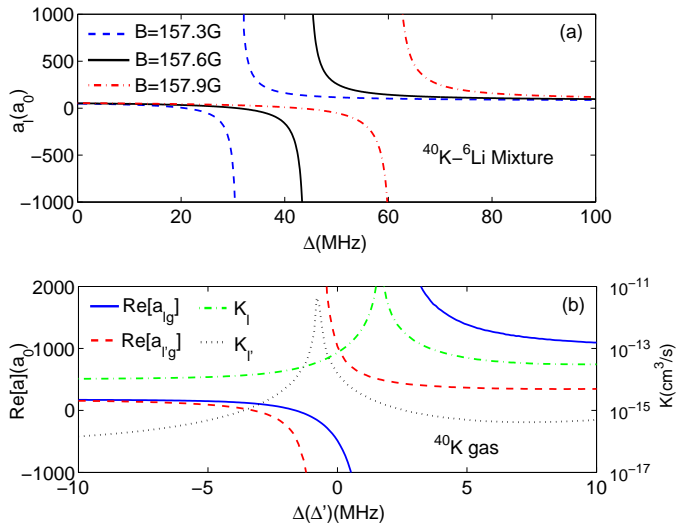}
\caption{(color online) (a): Control of two scattering lengths in
the three-component gases of a $^{\mathrm{40}}$K-$^{\mathrm{6}}$Li
mixture. The scattering length $a_{lg}$ is plotted in units of the
Bohr radius $a_0$ as a function of $\Delta$ with $\Omega=40$MHz and
$B=157.3$G (blue dashed curve), 157.6G (black solid curve) and
157.9G (red dash-dotted curve). (b): Control of two scattering
lengths in the three-component gases of $^{\mathrm{40}}$K with
$B=200$G and $\Omega=\Omega'=2$MHz. $\mathrm{Re}[a_{lg}]$ (blue
solid curve) and $K_l$ (green dash-dotted curve) are plotted as
functions of $\Delta$; $\mathrm{Re}[a_{l'g}]$ (red dashed curve) and
$K_{l'}$ (black dotted curve) are plotted as functions of $\Delta'$.
Since the scattering lengths between $|f_1(f_1')\rangle$ and
$|g\rangle$ are not available, we take the values to be the same as
those between $|f_2(f_2')\rangle$ and $|g\rangle$, which are given
in Refs. \cite{K7,K5,lik}. }
\end{figure}

The scattering and resonance between atoms in the dressed ground
states have been discussed in the literatures
\cite{rfscattering1,rfscattering3}. A new point
in our scheme is that we not only couple the two hyperfine states $%
|f_{1,2}\rangle$, but also employ the MFR between
$|f_{2}\rangle|g\rangle$ and $|c\rangle$. As shown above, the
enhancement of $a_{lg}$ usually occurs
when the dressed channel $|f_l\rangle|g\rangle$ is in the resonance region of $%
|\phi_{\mathrm{res}})$. Therefore with the help of the MFR, we can
control the location of the resonance of $a_{lg}$ by varying both
the dressing parameters and the magnetic field.

\textit{Independent control of two scattering lengths.} There are
two methods to realize the independent control of two scattering
lengths in a three-component system with our approach. The first
method (Fig. 1c) is to use the states $|g\rangle $, $|f_{1,2}\rangle
$, and an additional
hyperfine state $|d\rangle $. As shown above, the two states $%
|f_{1,2}\rangle $ are coupled and form two dressed states $|f_{h,l}\rangle $%
. We further assume the scattering length $a_{dg}$ between $%
|g\rangle $ and $|d\rangle $ can be tuned via a MFR which is close
to the one between $|2\rangle $ and $|c\rangle$. Therefore, in the
three-component system with fermionic atoms in the states
($|g\rangle ,|d\rangle
,|f_{l}\rangle $) we can control $a_{dg}$ by changing the $B$%
-field. Once the magnetic field is tuned to an appropriate value, we
can control the scattering length $a_{lg}$ by changing the coupling
parameters $\left( \Omega ,\Delta \right) $ with the approach
described above.

In the second method, we make use of five hyperfine states as shown
in Fig. 1e. We assume the static magnetic field is tuned to an
appropriate value so that the bare channel $|2\rangle $ is near MFR
with a bound state in a closed channel $|c\rangle $, while
$|2'\rangle\equiv|f_{2}^{\prime }\rangle
|g\rangle $ is also in the region of MFR with another closed channel $%
|c^{\prime }\rangle $. We assume that two states $|1,2\rangle $
are coupled with Rabi frequency $\Omega $ and detuning $\Delta $,
and form two dressed states $|f_{h,l}\rangle $. Similarly the
states $|f_{1,2}^{\prime
}\rangle $ are coupled with Rabi frequency $\Omega ^{\prime }$ and detuning $%
\Delta ^{\prime }$, and form dressed states $|f_{h,l}^{\prime }\rangle $.
Then according to the above discussion, the scattering length $a_{lg}$
between $|f_{l}\rangle $ and $|g\rangle $ can be resonantly controlled by ($%
\Omega $, $\Delta $), while $a_{l^{\prime }g}$ between $|f_{l}^{\prime
}\rangle $ and $|g\rangle $ can be resonantly controlled by ($\Omega
^{\prime }$, $\Delta ^{\prime }$). Therefore in the three-component system
with fermionic atoms in the states ($|g\rangle ,|f_{l}\rangle
,|f_{l}^{\prime }\rangle $), one can independently control $a_{lg}$ and $%
a_{l^{\prime }g}$ by changing the four parameters ($\Omega ,\Delta ,\Omega
^{\prime },\Delta ^{\prime }$).

In the above two methods, two conditions are required to make the
independent resonance controls of $a_{lg}$ and $a_{dg}$
($a_{l^{\prime }g}$) practical. First, the two MFRs for the bare
channels $|2\rangle $ and
$|d\rangle |g\rangle $ ($|2^{\prime }\rangle |g\rangle $%
) should be close to each other (e.g., the distance between the two
resonance points $\lesssim $ 10G). Second, Rabi frequencies $\Omega $ and $%
\Omega ^{\prime }$ should be large enough (on the order of $10$MHz).
Fortunately we can find such resonances in many systems \cite%
{K7,K5,lik,KRb}. In principle, the second method can be generalized
to the independent control of $n-1$ scattering lengths in an
$n$-component system. To this end, one should make use of $n-1$\
MFRs which are close together.

\textit{Possible experimental realizations.}  Now we discuss
possible implementations of our method. In a mixture of
$^{\mathrm{40}}$K and $^{\mathrm{6}}$Li \cite{lik}, we can realize
the three-level system with the first method in the above section.
To this end, we use the hyperfine state $|1/2,1/2\rangle $ of
$^{\mathrm{6}}$Li to be $|g\rangle $ and
consider the state $|9/2,-9/2\rangle $ of $^{\mathrm{40}}$K to be $%
|d\rangle $. We also take the states $|9/2,-7/2\rangle $ and $%
|9/2,-5/2\rangle $ of $^{\mathrm{40}}$K as $|f_{2}\rangle $ and $%
|f_{1}\rangle $, and form the dressed states $|f_l\rangle $. With
the help of the MFR in the channel $|d\rangle|g\rangle$ with
$B_{0}=157.6$G, $\Delta
B=0.15$G \cite{lik} and the one of $|2\rangle $ with $%
B_{0}=159.5$G, $\Delta B=0.45$G \cite{lik}, one can tune $a_{dg}$ by
changing the magnetic field, and then tune $a_{lg}$ by changing the
coupling parameters ($\Omega $, $\Delta $). We note that, since both
of the two bare channels $|1\rangle $ and $|2\rangle $ are stable,
there is no hyperfine relaxation in the collisions between
atoms in the states $|g\rangle $ and $|f_l\rangle $. The unequal masses of $^{%
\mathrm{40}}$K and $^{\mathrm{6}}$Li can lead to many new phenomena
in such a three-component heteronuclear system ($^{\mathrm{40}}$K
atoms in $|d\rangle $, $|f_l\rangle $ and $^{\mathrm{6}}$Li atoms in $%
|g\rangle $), e.g., the appearance of an Efimov state formed by two
heavy atoms and one light atom when both $a_{dg}$ and $a_{lg}$ are
large enough \cite{efimov}.

With a multi-channel calculation based on the realistic
$^{\mathrm{40}}$K-$^{\mathrm{6}}$Li interaction potential, one can
determine the scattering length $a_{lg}$ as a function of ($\Omega $, $%
\Delta $). For simplicity, we use a square-well model
\cite{squarewell}
for the $^{\mathrm{40}}$K-$^{\mathrm{6}}$Li interaction: $%
V_{i}^{(\mathrm{bg)}}(r)=-v_{i}^{(\mathrm{bg)}}\theta (\bar{a}-r)$, $%
W(r)=w\theta (\bar{a}-r)$ and $V_{1a}(r)=0.$ where $\theta(x)$ is
the unit step function and $\bar{a}$ is defined as
$\bar{a}=4\pi\Gamma(1/4)^{-2}R_{\rm vdW}$ with $\Gamma(x)$ being the
Gamma function and $R_{\rm vdW}$ the van der Waals length
\cite{squarewell}. $v_{i}^{(\mathrm{bg)}}$, $w$, $v_{1a}$ and
$\delta$ are determined by the experimental scattering parameters.
In the absence of the coupling between $|f_{1}\rangle $ and
$|g\rangle $, our model gives the background scattering lengths in
the channels $|f_{1}\rangle |g\rangle $ and $|f_{2}\rangle |g\rangle
$, and the correct resonance point and width for the MFR between
$|f_{2}\rangle |g\rangle $ and $|c\rangle $. Although the
square-well model cannot provide a quantitatively accurate result
for $a_{lg}$, it can give a qualitative and intuitive illustration
of our method. The results of our calculation are shown in Fig. 2a.
It is clear that, for every given value of the magnetic field (or
the scattering length $a_{dg}$), $a_{lg}$ can be resonantly
controlled via the laser induced coupling ($\Omega $, $\Delta $).

It is also possible to implement our method in ultracold $^{\mathrm{%
40}}$K atoms with the second method described above. We can utilize
five hyperfine states to construct the five-level structure as
schematically illustrated in Fig. 1e. The ground hyperfine state
$|9/2,-9/2\rangle $ is
used for $|g\rangle $, while four other states $|7/2,-5/2\rangle $, $%
|9/2,-7/2\rangle $, $|9/2,-3/2\rangle $ and $|9/2,-5/2\rangle $,
respectively are taken
as $|f_{1}\rangle $, $|f_{2}\rangle $, $|f_{1}^{\prime }\rangle $ and $%
|f_{2}^{\prime }\rangle $. To utilize the MFR in the channel $%
|2\rangle $ with $B_{0}=202.1$G, $\Delta B=8$G \cite{K7} and
the one in the channel $|2'\rangle $ with ($%
B_{0}=224.2$G, $\Delta B=10$G) \cite{K5}, we choose $%
B=200 $G. As shown above, dressed state $|f_l(f_{l^{\prime
}})\rangle $ can be
formed by the couplings between $|f_{1}(f_{1}^{\prime })\rangle $ and $%
|f_{2}(f_{2}^{\prime })\rangle $, and one can independently tune ($%
a_{lg},a_{l^{\prime }g}$) by choosing different dressing parameters
in the three-component gas of atoms in the states $(|g\rangle
,|f_l\rangle ,|f_{l^{\prime }}\rangle )$.

In this scheme, the upper hyperfine states $|f_{1}(f_{1}^{\prime
})\rangle $ can induce the HFR processes of both bare channels
$|1\rangle $, $|1'\rangle\equiv|f_{1}'\rangle$ and dressed channels
$|f_l(f_{l^{\prime }})\rangle |g\rangle
$. The values of the loss rate of the bare channels $%
|1(1')\rangle |g\rangle $ of $^{\mathrm{40}}$K are not available.
However, the loss rate of the process $|9/2,5/2\rangle
|9/2,7/2\rangle \rightarrow |9/2,9/2\rangle |9/2,3/2\rangle $ has
been measured to be as low as $10^{-14}{\rm cm^{3}/s}$
\cite{Krelaxation}. We take this value to be the loss rate of
$|1(1')\rangle$ in our calculation with the square-well model, where
$V_{1a(1'a')}(r)$ is assumed to be $v_{1a(1'a')}\theta(\bar{a}-r)$.
Both intra-channel scattering lengths (${\rm Re}[a_{lg}], {\rm
Re}[a_{l^{\prime }g}]$) and the two-body loss rate ($K_l,K_l^{\prime
}$) of $|f_l(f_l^{\prime })\rangle |g\rangle $ are shown in Figs. 2b
and 2c. We find that, around the resonance point where the loss
rates ($K_l,K_l^{\prime }$) peak, there are broad regions with large
absolute values of (${\rm Re}[a_{lg}], {\rm Re}[a_{l^{\prime }g}]$)
and small ($K_l,K_l^{\prime }$). Although we do not perform the
calculation with realistic interactions potential between $^{40}$K
atoms, our result based on the square-well model also shows the
feasibility of our scheme in a gas of $^{\mathrm{40}}$K.

\textit{Conclusion and discussion.} In this Letter we propose a
method for the independent control of (at least) two scattering
lengths in three-component gases by preparing atoms in dressed
states and via independent tuning of the couplings among the
hyperfine states. Under suitable conditions, our method can be
generalized to the control of $n-1$ ($n>3$) scattering lengths in an
$n$-component system. This would be a powerful technique for
engineering different types of homonuclear or heternuclear Efimov
states and for control of quantum phases. We have shown that our
scheme can be implemented in cold gases of $^{\mathrm{40}}$K or a $^{\mathrm{%
40}}$K-$^{\mathrm{6}}$Li mixture. It is also possible to apply our method to
bosonic systems, such as the $^{\mathrm{40}}$K-$^{\mathrm{87}}$Rb mixture,
where two close Feshbach resonances at $B$=464G and 467.8G \cite{KRb} are
available.

We thank M. A. Cazalilla, S. Inouye, S. Jochim, Y. Kawaguchi, T.
Mukaiyama, T. Shi, C. P. Sun, Y. Takahashi and R. Zhao for helpful
discussions. We also gratefully acknowledge
thank E. Tiesinga for providing the interaction potentials of $^{\mathrm{6}}$%
Li atoms.

\end{document}